\documentstyle{mn2e}
\input epsf

\def\deg{\hbox{$^\circ$}}

\def\lesssim{\mathrel{\hbox{\rlap{\hbox{\lower4pt\hbox{$\sim$}}}\hbox{$<$}}}}
\def\gtrsim{\mathrel{\hbox{\rlap{\hbox{\lower4pt\hbox{$\sim$}}}\hbox{$>$}}}}

\begin{document}

\title[IPHAS A stars]
{Early-A stars from IPHAS, and their distribution in and around the  Cyg OB2 
association}
\author[Janet E. Drew et al. ]
{Janet E. Drew$^{1,2}$, R. Greimel$^{3,4}$, M. J. Irwin$^5$, S. E. Sale$^1$\\  
$^1$Physics Department, Imperial College London, 
Exhibition Road, London,  SW7 2AZ, U.K.\\
$^2$Centre for Astrophysics Research, STRI, University of Hertfordshire,
College Lane Campus, Hatfield, AL10 9AB, U.K.\\
$^3$Institut f\"ur Physik, Karl-Franzens Universit\"at Graz,
Universit\"atsplatz 5, 8010 Graz, Austria\\
$^4$Isaac Newton Group of Telescopes, Apartado de correos 321,
E-38700 Santa Cruz de la Palma, Tenerife, Spain\\
$^5$Institute of Astronomy, Madingley Road, Cambridge CB3~0HA, U.K.
\\
}

\date{received,  accepted}

\maketitle

\begin{abstract}
Stellar photometry derived from the INT/WFC Photometric H$\alpha$ Survey of
the Northern Galactic Plane (IPHAS) can be used to identify large, reliable
samples of A0-A5 dwarfs. For every A star, so identified, it is also possible 
to derive individual reddening and distance estimates, under the assumption
that most selected objects are on or near the main sequence, at a mean 
absolute $r'$ magnitude of 1.5 -- 1.6.  This study presents the method for 
obtaining such samples and shows that the known reddenings and distances
to the open clusters NGC 7510 and NGC 7790 are successfully recovered.  
A sample of over 1000 A stars is then obtained from IPHAS data in the 
magnitude range $13.5 < r' < 20$ from the
region of sky including the massive northern OB association Cyg~OB2.
Analysis of these data reveals a concentration of $\sim$200 A stars over an 
area about a degree across, offset mainly to the south of the known 
1--3~Myr old OB stars in Cyg~OB2: their dereddened $r'$ magnitudes 
fall in the range 11.8 to 12.5.  These are consistent with a $\sim7$~Myr 
old stellar population at $DM = 10.8$, or with an age of $\sim5$~Myr at 
$DM = 11.2$.  The number of A stars found in this clustering alone is
consistent with a lower limit to the cluster mass of $\sim 10^4$~M$_{\odot}$. 
\end{abstract}


\section{Introduction}

   In the spectra of stars, the H$\alpha$ line is a prominent marker both as 
an emission line and as an absorption feature.  H$\alpha$ surveys are generally
seen as vehicles for identifying emission line nebulae and stars associated 
with, primarily, the early and late phases of stellar evolution.  In the case
of IPHAS, the INT/WFC Photometric H$\alpha$ Survey of the Northern Galactic
Plane (Drew et al. 2005), the quality of the digital photometry in the 3 survey 
filters, $r'$, $i'$ and $H\alpha$, also allows the strong H$\alpha$ absorption 
associated with A-type stars near or on the main sequence to be used as a tool 
to select these early-type objects easily and reliably.  Furthermore, the 
photometry is sufficient to provide estimates of reddenings and distances.  
The aim of this paper is to demonstrate how this capability may be exploited.
In so doing, we shall verify the method against two test cases, and present 
an example of this selection at work.

   We shall show that the selection of A stars from the IPHAS $(r'-H\alpha, 
r'-i')$ colour-colour plane yields mainly samples of A0-A5 dwarf candidates.
These early-type dwarfs of intermediate mass are expected to remain near the 
main sequence for not much longer than $\sim$100 million years, and so will 
typically trace relatively young environments -- which are a defining feature 
of the thin disk population of our Galaxy.  Objects of this type selected via
their maximal H$\alpha$ absorption have been exploited before as probes of 
structure within the galactic plane: in a series of papers, S. McCuskey and 
collaborators identified them in objective prism data down to $V \sim 12$ and 
then mapped their distribution (see e.g. McCuskey \& Houk 1971, McCuskey \& 
Lee 1976).  This technique can be picked up again and applied to the IPHAS 
database starting from a bright limit of $\sim 13$th magnitude, reaching 
down to $\sim$20th.  As IPHAS approaches completion, and its catalogue
database is becoming public (Gonzalez-Solares et al. 2008), the targeted 
selection of A stars can be applied to a range of problems across the northern 
Galactic plane, ranging from focused studies of individual clusters to
tracing large scale Galactic disk structure.

    The main application considered here is of the more localised type -- to 
the still enigmatic Cyg~OB2 association: data from 21 IPHAS fields covering 
the sky in and around this massive OB association are presented and analysed.
Cyg~OB2 is the most impressive of the northern hemisphere's Galactic OB 
associations.  Our aim is to use the A-star distribution to better define
the extent and nature of this massive cluster.  This example also serves to 
show that even very young regions -- 1--3~Myrs old in this case (Hanson 2003) 
-- can be usefully explored by this means.

    Cyg~OB2 (VI Cygni) is home to the only known northern hemisphere O3 stars
(Walborn et al. 2002).  It was first recognised for the very massive cluster 
that it is by Reddish, Lawrence \& Pratt (1967), who carried out the first 
extensive study of the region.  They placed it at a distance of 2.1~kpc, 
based on the apparent magnitude of late B stars.  They also defined its 
extent as a 29~pc $\times$ 17~pc ellipse centred on RA 20 32 30 
Dec +41 30 00 (J2000), and placed its mass in the range 0.6 -- 
2.7$\times 10^4$~M$_{\odot}$.  However now, mainly as the result of the work 
by Massey \& Thompson (1991), the main concentration of OB stars is recognised 
as lying just to the NE of the VI~Cyg~No 8 trapezium located at RA 20 33 06 
Dec +41 22 30 (J2000), at the nearer distance of 1.7~kpc
($DM = 11.2$).  This was not so clearly evident, to begin with, because of the 
substantial reddening of the main association: $A_V \sim 6$ is typical for 
Massey \& Thompsons's (1991) OB population.  On the basis of 2MASS NIR 
photometry 
of the region, Kn\"odlseder (2000) argued that Cyg~OB2 could be even more 
massive than Reddish et al's (1967) estimate, perhaps reaching as much as 
$10^5$~M$_{\odot}$, and could contain as many as 2600$\pm$400 OB stars.  
This has since been challenged by Hanson (2003), who has also shown that 
recent recalibrations of OB star absolute magnitudes may require a further 
reduction of the distance to Cyg~OB2 to $\sim$1.4~kpc (or $DM = 10.8$). 

    The paper is organised as follows.  The next section presents the main
features of IPHAS and the resultant database.  Then in Section 3, the IPHAS
$(r'-H\alpha, r'-i')$ colour-colour plane is discussed from the perspective
of the reliable selection of A stars from it, how they may be dereddened, and 
how this selection maps onto a mean absolute magnitude.  We then present two 
test cases, in the form of the well-studied open clusters NGC 7510 and 
NGC 7790, to show that the selection works as expected in recovering their 
known distances and reddenings.  This prepares the way for the
challenge of exploring the Cyg~OB2 region in section 5, in which the full 
impact of the early A star absolute magnitude sensitivity to age, at the
youngest ages, is seen.  We conclude with a comment on our findings and
on further development and application of the A-star selection technique.

\section{Observations: IPHAS photometry}

\begin{table*}
\caption{Co-ordinates of field-pair centres, extracted sky areas and object 
counts.} 
\label{field_list}
\begin{tabular}{lllcccrrr}
\hline
IPHAS & \multicolumn{4}{c}{Extraction Centre Coordinates} & Extracted area &
   \multicolumn{3}{c}{Star counts: $13.5 \leq r' \leq r'_{\rm max}$} \\
field & RA & Dec & $\ell$ & $b$ & & total & \multicolumn{2}{c}{candidate A dwarfs} \\
number & (2000) & (2000) & ($\deg $) & ($\deg $) & (arcmin $\times$ arcmin) & &  (0.02 cut) & (0.03 cut) \\   
\hline
\hline
NGC 7510 & & & & & & \multicolumn{3}{c}{$r'_{\rm max}=19.0$} \\
7366 & 23 11 00 & $+$60 34 00 & 110.90 & $+$0.06 & 6$\times$6 & 460 & 34 & 53 \\
\hline
NGC 7790 & & & & & & \multicolumn{3}{c}{$r'_{\rm max}=19.5$} \\
7626 & 23 58 24 &   $+$61 13 00 & 116.60 & $-$1.00 & 6$\times$6 & 601 & 33 & 43 \\
7626 & 23 58 08 &   $+$61 24 30 & 116.60 & $-$0.81 & 31$\times$31 & 8623 & 68 & 140 \\
\hline  
\multicolumn{2}{l}{Cyg OB2 region} & & & & & \multicolumn{3}{c}{$r'_{\rm max}=20.0$} \\
5937 & 20 28 00 &   $+$42 20 30 & 80.48 & $+$2.18 & 31$\times$31 & 2037 & 41 & 57 \\
5939 & 20 28 12 &   $+$40 52 30 & 79.31 & $+$1.30 & `` & 2362 & 60 & 83 \\
5951 & 20 29 01.6 & $+$41 25 30 & 79.85 & $+$1.49 & `` & 2363 & 107 & 133 \\
5962 & 20 29 50.6 & $+$41 58 30 & 80.38 & $+$1.69 & `` & 2472 & 43 & 57 \\
5965 & 20 30 00.4 & $+$40 30 30 & 79.21 & $+$0.81 & `` & 1448 & 49 & 65 \\
5973 & 20 30 42 &   $+$42 31 30 & 80.92 & $+$1.89 & `` & 2251 & 57 & 80 \\
5976 & 20 30 51	&   $+$41 03 30 & 79.86 & $+$0.85 & `` & 2258 & 68 & 96 \\
5985 & 20 31 40.5 & $+$41 36 30 & 80.29 & $+$1.20 & `` & 4334 & 172 & 223 \\
5998 & 20 32 30.5 & $+$42 09 30 & 80.82 & $+$1.40 & `` & 2195 & 80 & 98\\
6001 & 20 32 39   & $+$40 41 30 & 79.66 & $+$0.51 & `` & 2067 & 46 & 67\\
6008 & 20 33 21.5 & $+$42 42 30 & 81.36 & $+$1.60 & `` & 1716 & 19 & 34\\
6010 & 20 33 30   & $+$41 14 30 & 80.19 & $+$0.71 & `` & 2942 & 48 & 77\\
6021 & 20 34 20.7 & $+$41 47 30 & 80.73 & $+$0.91 & `` & 2274 & 47 & 70\\
6023 & 20 34 27.5 & $+$40 19 30 & 79.57 & $+$0.01 & `` & 1090 & 4 & 8 \\  
6031 & 20 35 10.8 & $+$42 20 30 & 81.26 & $+$1.11 & `` & 1294 & 7 & 9 \\
6035 & 20 35 17   & $+$40 52 30 & 80.10 & $+$0.22 & `` & 2169 & 77 & 102 \\
6046 & 20 36 09.5 & $+$41 25 30 & 80.64 & $+$0.42 & `` & 1509 & 8 & 16 \\
6057 & 20 37 01.3 & $+$41 58 30 & 81.18 & $+$0.62 & `` &  951 & 0 & 3 \\
6058 & 20 37 07.5 & $+$40 30 30 & 80.02 & $-$0.28 & `` & 1018 & 5 & 7  \\
6065 & 20 37 52.4 & $+$42 31 30 & 81.71 & $+$0.83 & `` & 1221 & 5 & 12 \\
6067 & 20 37 58   & $+$41 03 30 & 80.55 & $-$0.08 & `` & 1151 & 8 & 10 \\
\hline
   &   &   &   & \multicolumn{3}{r}{(Cyg OB2 region totals:)} & 910 & 1250 \\
\hline
\hline 
\end{tabular}
\end{table*}
     
    A full description of IPHAS data-taking and data-processing has been given 
by Drew et al. (2005).  Here, we mention again only the key features of the 
observations.  

     All IPHAS imaging is obtained with the Wide Field Camera (WFC), mounted 
on the 2.5-metre Isaac Newton Telescope, in La Palma.  The WFC is an imager 
comprising 4 AR-coated, thinned 4K $\times$ 2K EEV CCDs arranged in an L shape,
capturing data from an on-sky area of approximately 0.3 of a square degree.  
With a pixel dimension of 13.5~$\mu$m, corresponding on-sky to 
0.333$\times$0.333 arcsec$^2$, the instrument is able to exploit the 
sub-arcsecond seeing frequently encountered at the telescope site.  All 
the observations of the Cyg OB2 region, presented here, were obtained 
during the clear moonless nights of 8 and 9 August 2004.  The data on the 
open clusters NGC 7510 and 7790 were obtained on grey nights, in stable
conditions, respectively on 26 July and 22 October 2005.  

\begin{figure}
\begin{picture}(0,250)
\put(0,0)
{\includegraphics{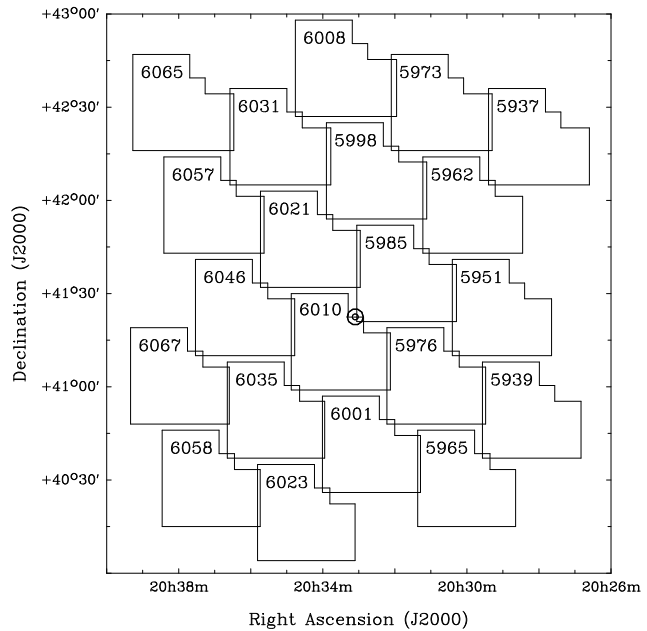}}
\end{picture}
\caption{The positions of the extracted IPHAS field-pair overlaps in
and around CygOB2.  The black concentric circles in between fields
5985 and 6010 mark the often-quoted centre of Cyg OB2 (e.g.
as in Kn\"odlseder 2000). 
}
\label{figure_fields}
\end{figure}

  A complete IPHAS observation of any sky position consists of two pointings,
the second at an offset of 5 arcmin W and 5 arcmin S with respect to the 
first.  This strategy both enhances the overall photometric quality of the 
final survey and ensures that stars falling into a gap between the mosaiced 
CCDs in one exposure are captured in the partner exposure.  In this study
we only consider point sources that are well-detected in both exposures: this
limits the sky extent of the overlap between each field pair to a near-square 
of side 31 arcmin.  Within this there is an irregular grid of inter-CCD gaps, 
with the result that the sample of point sources discussed here drawn from 
each field pair comes from a sky area of 0.22 deg$^2$: since there is very 
little overlap between adjacent field pairs when they are combined in this 
way, all the Cyg~OB2 fields together span close to 90 percent of the total 
available area of $\sim5$~deg$^2$.  The rough sky outlines of the IPHAS 
field-pair overlaps in the Cyg~OB2 region are shown in 
Fig.~\ref{figure_fields}.  The coordinates of the centres of all catalogue 
extractions discussed in this paper are listed in Table~\ref{field_list}, 
along with other relevant data.  

  Throughout this paper field pairs are named just by the 4-digit number 
(column 1 in Table~\ref{field_list}) that tags the initial field pointing in 
the IPHAS database. 

    At each WFC pointing, imaging is carried out through three filters. 
These filters are narrow-band H$\alpha$ (FWHM 95~\AA ), Sloan $r'$ and 
Sloan $i'$.  The exposure times in these three filters are 120~sec 
($H\alpha$), 30~sec ($r'$) and 10~sec ($i'$).  Since it is required that 
every star considered is imaged in both overlapping pointings, the effective 
exposure times are twice as long.  Every night of IPHAS observing 
incorporates observations of standard fields, in order to permit photometric 
calibration in the three observed bands.  Presently the H$\alpha$ narrow-band
observations are tied to the $r'$-band calibration in the same way as 
described by Gonzalez-Solares et al. (2008).

    A further criterion applied in our selection of candidate A stars is 
that the CASU pipeline processing and cataloging the data, identifies them as 
point-like in $r'$ and $i'$, to a high confidence level (i.e. as members of
morphology classes -1 and -2, Irwin 1985 and 1997).  The photometric error in 
the resultant catalogue of point sources ranges from entirely negligible at 
the adopted bright limit (13th magnitude in $r'$), to, typically, 0.03 in 
$r'$, 0.06 in $(r' - H\alpha)$ and 0.04 in $(r' - i')$ at the chosen faint 
limit ($r' = 20$).

\section{Method}

\subsection{The IPHAS colour-colour plane, illustrated by an example from 
Cyg OB2}

\begin{figure*}
\begin{picture}(0,300)
\put(0,0)
{\includegraphics{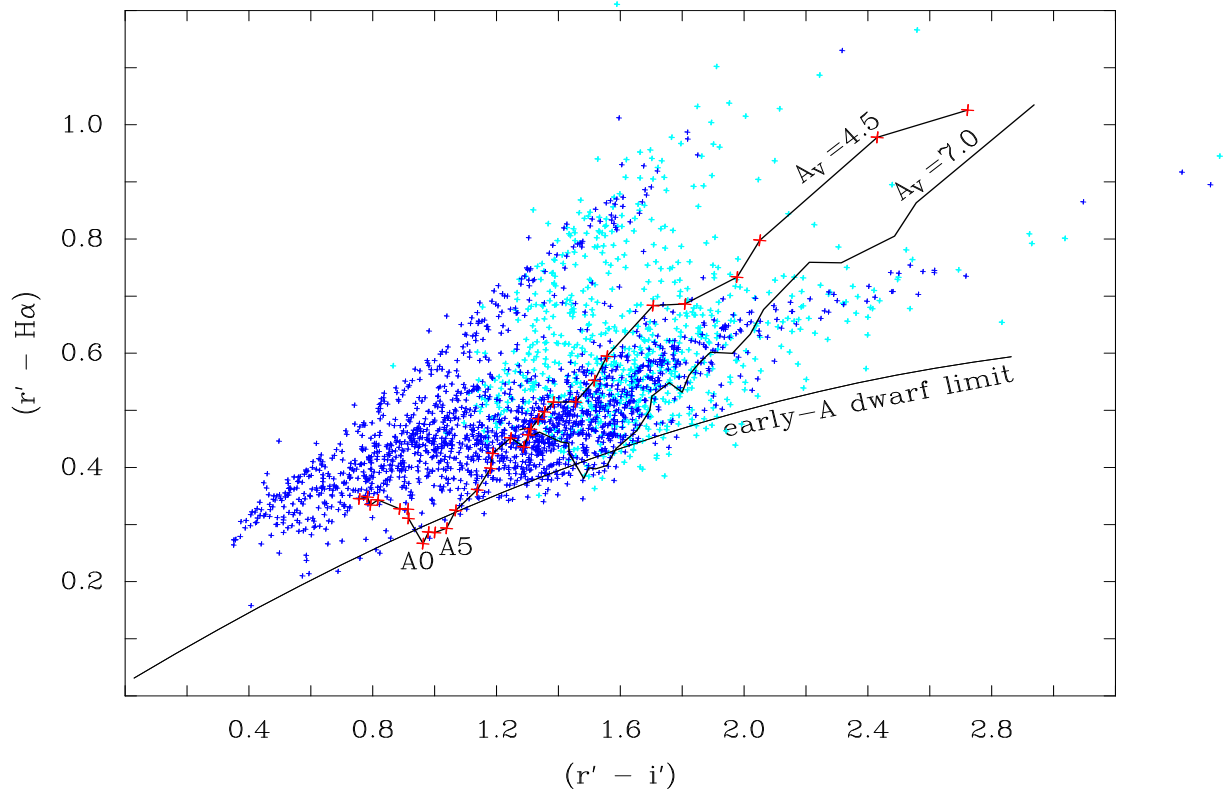}}
\end{picture}
\caption{The $(r' - H\alpha, r' - i')$ colour-colour plane of IPHAS field 6010
point sources, selected from the magnitude range $13 \leq r' \leq 20$.  The
black line running along below most of the data points and labelled 'Early A 
dwarf limit' is the reddening line below which the great majority of point 
sources are expected to be near mains sequence A0-A5 stars. Two reddened main 
sequences constructed from synthetic photometry based on the Pickles (1998) 
library of flux-calibrated spectra are also drawn (in black also).  The red 
crosses on the $A_V = 4.5$ sequence pick out the data points that make it up 
(spanning spectral types O5V to M4V). The IPHAS photometric data are 
colour-coded according to source $r'$ magnitude: $13 \leq r' < 19$, darker 
blue; $19 \leq r' < 20$, light blue.
}  
\label{cc_data}
\end{figure*}

    An example of derived colours for a final selection of point sources 
is shown in Fig.~\ref{cc_data}.  These are the data for IPHAS field 6010,
which includes the nominal centre of Cyg~OB2.  All 2943 stars plotted fall 
in the magnitude range $13 \leq r' \leq 20$.  To understand what is implied 
by this source distribution we make use of the solar-metallicity synthetic 
tracks that were given in Drew et al. (2005).  On Fig.~\ref{cc_data} we
have superimposed a curve that defines an approximate upper bound
on the colour-colour plane location of early A dwarf stars.  More precisely
this is the reddening line derived using as template the solar metallicity 
A0V flux-calibrated spectrum from the library due to Pickles (1998), shifted
to higher $r'-H\alpha$ by 0.03.  The reddening law used here is a mean 
Galactic law, with $R = 3.1$ (Howarth 1983).  

   The main features of the distribution of stars in the colour-colour plane 
(Fig.~\ref{cc_data}) are (i) a foreground unreddened main sequence,
beginning at a spectral type of $\sim$G0 (at $r' = 13$); (ii) an intermediate 
group of lightly reddened stars ($A_V \lesssim 3$), for which the earliest 
detected spectral type shifts to $\sim$F0 (at $r' \sim 14.5$); (iii) a 
concentration of stars spread between the reddened $A_V = 4.5$ and $A_V = 7$ 
MS lines, and reaching down to and below the early-A dwarf limit line.  It 
can be deduced from the observed magnitudes and colours that the stars in the 
first two groups will be located at distances of no more than $\sim$500~pc.  
This impression is consistent with a study by McCuskey and Houk (1971) of 
A stars brighter than $V = 13$ in which they remarked on a declining stellar 
density between 500~pc and 1~kpc in this part of the Plane, beyond a high 
concentration found between 200 and 300~pc (see Fig. 7 of their paper).
Group (iii) in Fig.~\ref{cc_data} is representative of the Cyg OB2 region.  
Previous work on Cyg OB2 itself have claimed typical 
reddenings in the range from $A_V \simeq 4$ to $\simeq 7$ (Massey \&Thompson
1991, Hanson 2003).  In Fig.~\ref{cc_data} it can be seen that the 
colour-colour plane is well occupied in this range.  It is worthy of note that 
redward of the $A_V = 7$ main sequence there is a distinct thinning 
out of stars.  This is mainly attributable to the faint limit imposed
on the included data (cf. Fig.~\ref{fig_DMvred} discussed in section 5). 

\subsection{The ingredients of IPHAS early-A star selection}

At the heart of the method of selection is the choice of a suitable cut 
along the lower edge of the main stellar locus in the ($r'-H\alpha,r'-i'$) 
plane that separates off candidate early-A dwarfs from what will most 
commonly be later-type stars.  Once a group of candidate objects has been 
picked out in this way, it is straightforward to use the measured $r'-i'$ of 
each object to derive its reddening, and then to deredden its measured 
apparent $r'$ magnitude.  If a suitable averaged absolute magnitude is 
available, it becomes possible as a next step to determine the distance 
modulus ($DM = 5\log D - 5$) for the sample.  This enables the construction 
of a reddening map for the area of sky studied and gives a picture of how 
the A stars are distributed with distance.  From this approach, one does not 
expect to derive accurate reddenings and distances for individual candidate
A dwarfs, for a number of reasons including that the photometry cannot pin 
down the A sub-type to within better than 2 to 3 subtypes, and that the
absolute magnitude spread at each sub-type is in any case significant
(Jaschek \& Gomez 1998).  What should emerge is a reliable picture in a 
statistical sense, insofar as the mean properties of the selected population 
of objects are well-defined.

The shape of the lower edge of the main stellar locus in the ($r'-H\alpha,
r'-i'$) plane will always follow a segment of the reddening curve of an 
early-A dwarf, as was first noted by Drew et al. (2005).  Such a curve is
therefore appropriate to use as the cut line.  Furthermore, in 
Fig.~\ref{cc_data},
it can be seen from the labelling of the $A_V = 4.5$ main sequence that
spectral types A0 to A5 form a line segment running essentially parallel
to the reddening line labelled ``A dwarf limit'' (actually the A0V reddening
line shifted upward by $\Delta(r'-H\alpha) = 0.03$, in this instance).  
This is a direct consequence of the fact that the H$\alpha$ absorption 
equivalent width reaches a crudely defined maximum of 11 -- 13 \AA , 
across this dwarf sub-type range (cf. comment on this in Drew et al. 2005, 
section 5).  

\begin{figure}
\begin{picture}(0,350)
\put(0,0)
{\includegraphics{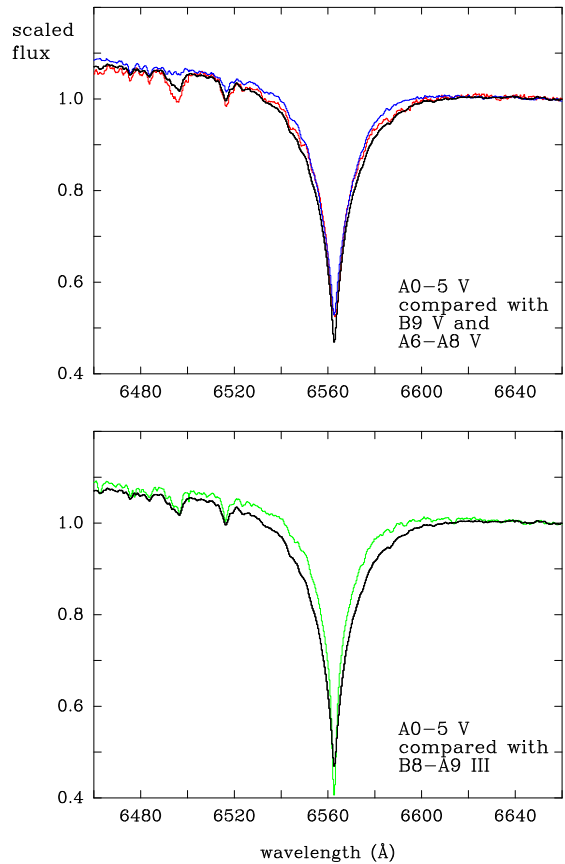}}
\end{picture}
\caption{
H$\alpha$ line profiles for spectral types of early-A
dwarf and adjacent spectral classes.  Each profile is the merger of 
many taken from the Indo-US library of coude spectra (Valdes et al. 2004).  
The top panel provides a comparison of A0-5 dwarfs (in black) with B9V 
(blue) and A6-8 (red) dwarfs, while the bottom shows the difference between 
A0-5 dwarfs and giants (green) at $\sim$A type. Before merging, the spectra 
were rescaled to yield unit flux at 6640~\AA , to facilitate the comparison
without introducing further corrections that might alter the very subtle 
depth differences across the central $\sim$80~\AA\ of the H$\alpha$ 
profile.   
}
\label{figure_ha}
\end{figure}

This is not just a quirk of the stars selected by Pickles (1998) that
were used as a basis for the synthetic photometry.  To confirm this
crucial point, the H$\alpha$ line profiles of near-solar metallicity A dwarfs 
extracted from the Indo-US spectral library (Valdes et al. 2004) have been 
compared with those of stars of neighbouring spectral type and luminosity 
class.  The results of this are summed up in simplified form in 
Fig.~\ref{figure_ha}.   The diagram shows that in the mean, A0-5 dwarfs
(data on 28 stars combined) are better separated from B9 dwarfs 
(17 stars), than from A6-A8 dwarfs (7 stars).  Given how spectral 
classification first arose, this is to be expected.  The luminosity class
comparison is not on such a strong statistical footing, as only 16 giant-star
spectra make up the comparison set.  Nevertheless, it is clear that the 
H$\alpha$ absorption equivalent width in an early-A giant is less, on
average, than in an early-A dwarf.  On an object by object basis there is
enough scatter that the distinction between, especially, $>$A6
and $<$A5 dwarfs is not always in the expected sense.  At the same time
there is some compensation through the empirical fact that the late A
near main sequence range is not as well populated with objects as at
early A: the bright sample of Abt \& Morrell (1995 Table 5) shows that
around 1 in 3 luminosity class V and IV stars are A5 or later.

\begin{figure}
\begin{picture}(0,150)
\put(0,0)
{\includegraphics{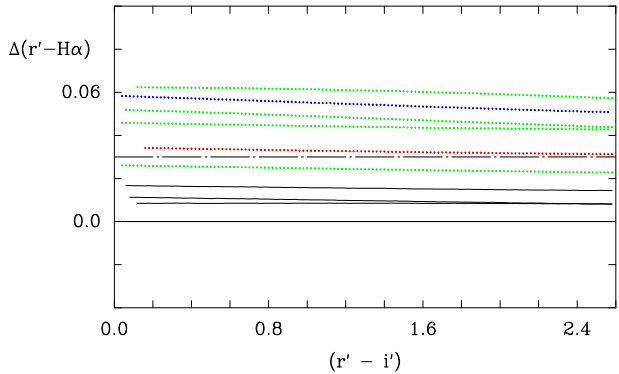}}
\end{picture}
\caption{
Reddening lines derived from the IPHAS colour-colour plane for 
spectral types in the early-A dwarf range plotted as differences with respect 
to the A0V reddening line.  The same colour code is in use as in 
Fig.~\ref{figure_ha}, e.g. A0-5 dwarf lines are drawn in black.
The fluxed spectra used to synthesise these tracks were taken from the
Pickles (1998) library.  The lowest line, along the horizontal axis, is the 
A0V reddening line.  Note that $\Delta(r' - H\alpha) = 0.02$ runs just below 
the A0~III line, the lowest green line; whereas  $\Delta(r'-H\alpha) = 0.03$, 
drawn in as a dash-dotted line, falls just under the A7V line (in red).  The
blue line is derived from the B9V reddening line. 
}
\label{figure_rlines}
\end{figure}

This pattern of behaviour translates into reddening lines in the 
$(r'-H\alpha,r'-i')$ plane as shown in Fig.~\ref{figure_rlines}. 
For clarity each reddening line drawn has had the A0V reddening line 
subtracted from it, causing the ordinate in the diagram to become
$\Delta(r' - H\alpha)$.  The A0-A5 V reddening lines, drawn in black, lie  
below those for adjacent dwarf spectral types and luminosity class III.
The appearance of this diagram gives an indication of the $\Delta(r'-H\alpha)$
range that can be associated exclusively with the early-A dwarfs: it is
likely to be $\sim$0.02, with the A0V curve used here defining 
$\Delta(r'-H\alpha) = 0.0$.   Photometric error, together with the 
inevitable gradual mixing in of (mainly) later type dwarfs as the cut line 
is raised, stands in the way of a fixed recommendation for placing of the 
cut line.  If the aim of selection is to be inclusive of all A dwarfs, 
choosing a cut at $\Delta(r'-H\alpha) = 0.04$ or 0.05 would make more sense.
Hales et al (2008) have obtained spectra of candidates selected using this
more generous criterion and find that they are the expected A stars.
 
Here we explore the effect of gradually raising the cut on the selection 
process, to see how increased sample size trades against increased likelihood 
of sample contamination.  The fit to the unshifted A0V line in the 
$(r'-H\alpha, r'-i')$ plane that we use is a simple quadratic:
\begin{equation}
    (r' - H\alpha) = -0.009 + 0.330 (r' - i') - 0.0455 (r' - i')^2
\end{equation}

The other quantitative issues to settle are the median intrinsic $r'-i'$
colour to associate with A0-A5 dwarfs (needed for dereddening), and  
the median absolute magnitude, $M(r')$ that should be adopted for them.  To
associate absolute visual magnitude with dwarf spectral type and colour, 
we use the data based on Hipparcos measurements presented by Houk et al. 
(1997).  To find suitable median values, it is necessary to consider how
a typical dwarf luminosity function would map onto the A0-A5 range and 
pick out the expected spectral type and colour of the median absolute 
magnitude for this spectral type range.  To do this, we use the main 
sequence luminosity function derived by Murray et al. (1997) also from 
Hipparcos data.  Houk et al. (1997) give $M_V = 0.82$ for A0, and 1.76 for 
A5 and note a spread of 0.39 mags in absolute magnitude at each spectral 
sub-type.  Based on the data in Murray et al. (1997), a median $M_V$ of 
$\sim$1.5 
is reasonable, for which the intrinsic $B-V$ is $\sim$0.1 (Houk et al. 
1997).  From the main sequence colours tabulated by Kenyon \& Hartmann 
(1995), intrinsic $V-R_C \simeq 0.03$, and $R_C-I_C \simeq 0.06$ are implied 
(spectral type A3-4).  

This line of reasoning leads to a median intrinsic $r'-i'$ of 0.06, for which 
the uncertainty is unlikely to be more than $\pm 0.02$.  To derive
the colour excess due to extinction, $E(B-V)$, from an observed value of 
$r'-i'$, this intrinsic colour is subtracted to give $E(r'-i')$, which is 
then multiplied by 1.58.  To obtain the visual extinction, $A_V$, $E(r' -i')$
should be multiplied by 4.91.  These transformations apply to the case of
$R = A_V/E(B-V) = 3.1$, defining the most commonly encountered Galactic 
extinction law.  Where non-standard reddening laws are thought likely to 
apply (e.g. $R$ approaching 5 as in sightlines dominated by molecular gas), 
it would make more sense to refer all reddenings to either the $r'$ or $i'$ 
band, rather than to $V$ in view of the fact that atypical laws vary away from 
the norm most strongly in the blue-to-visual part of the optical spectrum 
(Cardelli, Clayton \& Mathis 1989).  Specifically, the following relations
can be applied: $A_{r'} = 4.13 E(r'-i')$, and $A_{i'} = 3.13 E(r'-i')$
(using tabulated data from Schlegel, Finkbeiner \& Davis 1998).  In the 
cases examined here (sections~\ref{test_cases} and \ref{cygob2}), previous 
studies have found the average Galactic extinction law or laws very like 
it to be satisfactory.

The choice of representative absolute magnitude can be subject to 
systematic variation, given that the age spread of A dwarfs within 80~pc of 
the Sun (the samples used by both Houk et al. 1997, and Murray et al. 1997) is 
not guaranteed to be the same as that in another potential target region.  
Domingo \& Figueras (1999, Fig. 7) have analysed A3-9 main sequence stars 
drawn from essentially the same Hipparcos sample as that discussed by Houk et 
al (1997) and show that the earliest among them are indeed quite widely 
scattered in $M_V$, with mean magnitude $\sim$1.5.  But they also demonstrate 
these field stars lie on and smeared above the zero-age main sequence (ZAMS), 
expected to fall at $M_V \sim 2$ -- that is, these stars may have ages of up 
to $\sim$0.5~Gyr (based on comparison with isochrones due to Schaller et al. 
1992).  At ages greater than $\sim$1~Gyr all stars on or near the early A
main sequence will have evolved away. 

In Cyg OB2, a region that may be as young as 1--3 million years old 
(Hanson 2003), we can anticipate that stars with the intrinsic $(r'-i')$ 
colours of early-A dwarfs are more massive stars still approaching the ZAMS. 
Such objects are evolving almost horizontally on the HR diagram and are 
likely to arrive on the main sequence as mid-late B stars at an age of 
$\sim$5~Myrs (Siess, Dufour \& Forestini 2000).  Accordingly these are stars 
with significantly brighter $M(r')$ than those destined for the early-A ZAMS, 
after a pre main sequence evolution taking around 10~Myrs or so (Palla \& 
Stahler 1993; Siess et al. 2000).  

\begin{table}
\caption{The evolution of $M(r')$ for stars of early-A $(r'-i')$ with 
cluster age.  These rough figures are derived from the $Z = 0.02$ 
theoretical isochrones of Siess et al. (2000).  Note that at 
the youngest ages of $\sim$3~Myr or less, there is a greatly increased 
likelihood that these objects will exhibit circumstellar H$\alpha$ emission, 
rather than maximal absorption (i.e. they will not lie along the bottom of 
the IPHAS main stellar locus).} 
\label{tab_Mr}
\begin{tabular}{cc}
\hline
age (Myr) & typical M(r')\\
\hline  
 2 & 0.0 \\
 3 & 0.5 \\
 5 & 1.0 \\
 7 & 1.4 \\
10 & 1.8 \\
20--100 & 1.9--2.0 \\
300 & 1.6 \\
\hline
\end{tabular}
\end{table}

Because this age effect can be significant, we give in Table~\ref{tab_Mr}
some representative values for $M(r')$ as a function of age that have been 
derived from the isochrones of Siess et al. (2000).  Up to $\sim$10 Myrs,
$M(r')$ declines markedly, and then remains close to $\sim$2 for a
further $\sim$100 Myrs or so -- at the end of which, evolution away from the 
main sequence begins.  In the absence of any independent constraint on 
population age or mix, it seems likely that the best guess for the general 
field in the Galactic disk will be around or slightly fainter than the 
Hipparcos value of $M_V \simeq 1.5$, or $M(r')$ of 1.5 -- 1.6.  Since the 
age-dependence has obvious and serious ramifications for using the selection 
of early-A dwarfs to analyse Galactic disk sightlines,
we begin with two control populations to see how well the method works.  Our 
guinea pigs are the two open clusters, NGC 7510 and NGC 7790, aged 6-10 Myrs 
(Piskunov et al. 2004) and $\sim$100 Myrs (Bragalia \& Tosi 2006), 
respectively.

Finally, it is worth recalling that binarity will also have some effect on 
the average absolute magnitude of a population of early A stars.  We make
no explicit correction for this here beyond favouring $M(r') = 1.5$, since
we estimate it amounts to a brightening of the mean in the region of 0.1 
magnitudes.  The following considerations suggest this figure: roughly a half 
of all stars are binary systems; the mass ratio distribution is not far from 
flat (Halbwachs et al 2003); hence only in 20--30 \% of all binaries is the 
combined magnitude likely to brighten by more than 0.5 (see figure 1 in Hurley 
\& Tout 1998).

\section{Test Cases}
\label{test_cases}

\subsection{Specifics of the extraction of candidate early-A dwarfs}

The steps in the extraction are very simple.  First, $r'-H\alpha$ (as a
function of $r'-i'$) is used to select stars as candidate early-A dwarfs.
Once selected, the measured $r'-i'$ is converted to a reddening, which then
allows the measured $r'$ to be corrected to a dereddened magnitude.  

There is no redundancy in our selection of candidate objects from IPHAS: we 
use 3 independent quantities to determine 3 properties.  Hence it is helpful 
to bring in e.g. near-infrared photometry to provide an independent check on
the selection.  For this, we turn to 2MASS $J$ magnitudes - where available.
Hence, as we select objects below an early-A dwarf limit line, we also 
collect the 2MASS $J$ magnitude along with the IPHAS magnitudes and check 
that the dereddened $r'-J$ colour falls in the early A-dwarf range (set 
generously to be $-0.1 < r'-J < 0.3$ for A0-5V, cf. Kenyon \& Hartmann 1995).  
If the dereddened $r'-J$ for any object falls outside this range, the object 
is dropped from consideration.  This weeding eliminates between a few percent
to 30 percent, depending on the field.  If no 2MASS $J$ magnitude is available 
or it is an upper limit only, we retain the candidate A-dwarf unverified, 
rather than throw it away.  If we were not to do this, our sample size would 
be noticeably reduced (by a further factor of up to 20--30\% in the worst 
cases, again depending on the field -- it is greater where the reddening is 
lower).  

In all cases, the bright limit is set at $r' = 13.5$ in order to avoid objects 
that might be saturated in $i'$ -- particularly in the reddened fields found
in the Cyg~OB2 region.  The faint limit is set at a value appropriate to the 
sensitivity limit for the IPHAS field under consideration.  Whilst exploring
this technique of A-dwarf selection, we treat the position of the cut line in
the colour-colour diagram as a variable.  In particular, we obtain samples of 
candidates for 4 positions of the cut line: for $\Delta (r'-H\alpha) = 0.015$, 
0.02, 0.025 and 0.03, where $\Delta (r'-H\alpha)$ is the upward dispacement of 
the A0V reddening line given in equation 1.  

\subsection{The open cluster NGC 7510}

\begin{figure}
\begin{picture}(0,150)
\put(0,0)
{\includegraphics{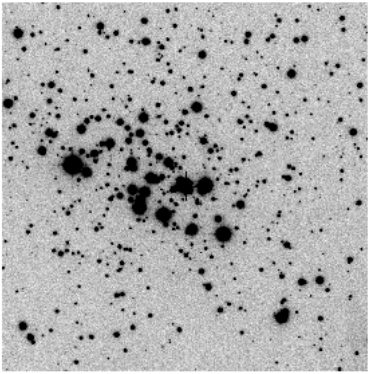}}
\end{picture}
\caption{The $6 \times 6$ arcmin$^2$ field centred on the test cluster 
NGC 7510, as extracted from the IPHAS database.  This is the 30-sec 
$r'$ exposures from pointing 7366o.  N is up and E to the left.
}  
\label{fig_7510r}
\end{figure}

\begin{figure}
\begin{picture}(0,190)
\put(0,0)
{\includegraphics{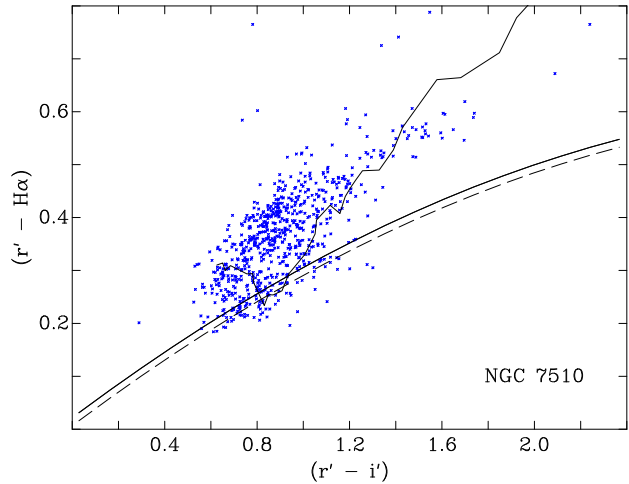}}
\end{picture}
\caption{The $(r' - H\alpha, r' - i')$ colour-colour plane of 
the $6\times6$ arcmin$^2$ field centred on the young open cluster NGC 7510
(in IPHAS fields 7366/o).  The point sources plotted are selected from the 
magnitude range $13.5 \leq r' \leq 19.0$.  The parallel solid and dashed lines 
are the early-A dwarf cut line shifted upward from the A0V reddening line by, 
respectively, 0.03 and 0.015.  The piecewise connected line is the 
$E(B-V)=1.2$ main sequence constructed from synthetic photometry based on 
the Pickles (1998) library of flux-calibrated spectra.  
}  
\label{fig_ngc7510}
\end{figure}

\begin{figure}
\begin{picture}(0,350)
\put(0,0)
{\includegraphics{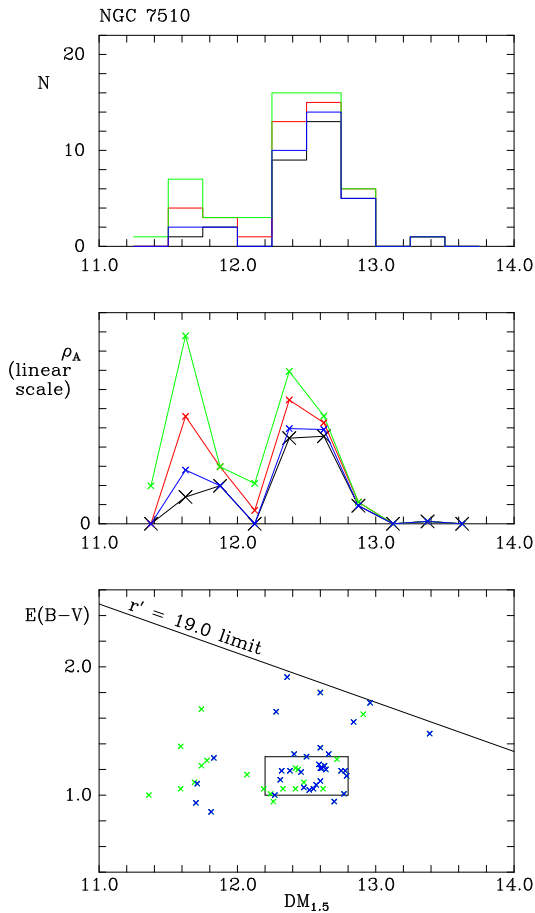}}
\end{picture}
\caption{The test case of the open cluster NGC 7510.  The upper panel
shows the distribution of derived distance moduli ($DM_{1.5} \equiv$ 
dereddened $r'$ magnitudes less an adopted median absolute magnitude of 
1.5).  The four curves shown are for four choices of positioning of the 
early-A dwarf limit line: $\Delta(r'-H\alpha) = 0.015$, 0.02, 0.025 and 0.03 
are respectively in black, blue, red and green.  Note that as the limit line 
is raised, in $\Delta (r'-H\alpha) = 0.005$ steps, the distribution is better 
populated but also begins to shift slightly toward lower mean distance 
modulus.  The middle
panel is a conversion of the upper panel to an A-star space density 
distribution, obtained by taking the number of stars in each histogram bin 
plot, and dividing by cube of the distance. The lower panel shows derived 
$E(B-V)$ versus $DM_{1.5}$, plotting first the largest sample obtained for 
the highest position of the cut line ($\Delta (r'-H\alpha) = 0.03$, in green), 
and then overplotting in blue the smaller sample obtained with the cut line 
lowered to $\Delta (r'-H\alpha) = 0.02$.  The box, centred on $DM=12.5$ and
$E(B-V)=1.15$, is the result for NGC 7510 due to Sagar \& Griffiths (1991).}
\label{fig_control}
\end{figure}

This young open cluster in the northern Galactic Plane has recently been 
studied by Piskunov et al. (2004), along with a number of southern 
examples.  The aim of this work was to examine the effect of relative 
youth (main sequence turn-on) on the cluster luminosity function.  
Piskunov et al. concluded that NGC 7510 is likely to be about 6 million
years old.  Other authors have favoured ages of $\sim$10 million 
years or more (Sagar \& Griffiths 1991, Barbon \& Hassan 1996, Kharchenko et
al 2005).    The CCD photometric study of the cluster by 
Sagar \& Griffiths (1991) earlier found the distance modulus to be 
12.5$\pm$0.3 and that $E(B-V)$ ranges from 1.0 to 1.3, in good agreement 
with e.g. Lynga's (1988) cluster catalogue.  

Our selection of candidate A dwarfs is taken from a $6\times6$ arcmin$^2$ 
region of sky centred on the cluster (Fig.~\ref{fig_7510r}).  The faint 
limit to the selection is
set at $r' = 19$ -- two magnitudes above the pipeline estimate of the 
magnitude limit for the observations of IPHAS fields 7366/o in which NGC 7510 
is located.  The lowest and highest cut lines applied to the data are shown 
in Fig.~\ref{fig_ngc7510}, superposed on the colour-colour diagram for 
NGC 7510 and environs obtained from the IPHAS database.

The results of the selection are shown in Fig.~\ref{fig_control}.  In the 
upper panel, the raw histogram of star numbers per $\Delta DM_{1.5} = 0.25$
bin is presented.  The distance modulus used here is derived applying a 
cluster age-appropriate value for $M(r')$ of 1.5 (see Table~\ref{tab_Mr}).
The extracted sample sizes are 31, 34, 43 and 53 objects, in order of 
increasing $\Delta(r'-H\alpha)$.  In the middle panel, an arbitrarily-scaled
quantity directly proportional to the space density of candidate
A-dwarfs, $\rho_A$, is plotted againt $DM_{1.5}$.  To arrive at $\rho_A$, the 
number of candidate A-dwarfs per distance modulus bin is divided by the cube 
of the bin mid distance: this corrects for the growing volume associated with 
each successive histogram bin.  The bottom panel shows the relationship 
between $DM_{1.5}$ and reddening for two samplings of the data.  

The main cluster is very clear, in all three panels of Fig.~\ref{fig_control}. 
In the histogram (upper panel), only 5 out of the sample of 34 stars for 
$\Delta (r'-H\alpha) = 0.02$ (blue line and points) fall outside the main 
peak, centred 
on $DM_{1.5} = 12.53\pm 0.04$ (measured by fitting a single Gaussian to the 
peak).  For the largest sample, obtained with $\Delta(r'-H\alpha) = 0.03$, 
12 stars out of 53 stray away from the cluster peak (green line and points), 
that itself has shifted downwards slightly to $DM_{1.5} = 12.48\pm0.02$.  The 
increased proportion of outliers hints that sample contamination may indeed be 
greater, but their impact is small.  Another property that does not change 
very much, as the cut line is raised, is the FWHM of the cluster peak: it 
remains close to $\Delta DM = 0.4$ in all 4 cases explored.  

As well as being clearly defined, the cluster is also at a mean
$DM_{1.5}$ and typical reddening that is in very good agreement with the
results of Sagar \& Griffiths (1991). The box superimposed on the 
reddening plot (bottom panel of Fig.~\ref{fig_control}) encloses the
reddening range they found, and the error bound on the distance 
modulus: $1.1 < E(B-V) < 1.3$ and $DM = 12.5\pm0.3$.   The 
extinction properties of the candidate A dwarfs reveal that there are 
objects -- within the cluster as far as we can tell -- at markedly higher 
extinctions ($E(B-V) > 1.5$) than found in prior optical studies.  

The outlier grouping at $DM_{1.5} \simeq 11.7$ that becomes more evident 
as the cut line is raised, appear to be as reddened as most of the main 
cluster group.  This similarity favours viewing these outliers as 
intrinsically brighter objects of similar optical colour within the 
cluster, rather than as members of an independent foreground cluster.
Indeed Piskunov et al. (2004) noticed that there was a case for a younger 
$\sim$2~Myr-old population in NGC 7510 in addition to the dominant 6--10 Myr
old component: the A dwarfs in such a component would be nearly 1 magnitude 
brighter as is implied here (e.g. the $M(r')$ difference between 3~Myr old
and 7~Myr old A stars is given in Table~\ref{tab_Mr} as 0.9 mags).  So we see 
that this selection technique begins to pick up on both generations of stars 
in NGC~7510.  

The middle panel of Fig.~\ref{fig_control}, corrected to trace A-star space 
density seemingly brings out this bimodal character even more.  However this 
more physical presentation of the data comes at a price: the correction to 
space density greatly amplifies both the count and the Poissonian error at 
lower $DM$ with respect to those at higher.  For example for the data for 
$\Delta (r'-H\alpha) = 0.03$ (coloured green), the error on the density peak 
in the range $11.5 < DM_{1.5} < 11.75$ is nearly 40 \% (just 7 stars), while 
it is under 20\% in the range $12.25 < DM_{1.5} < 12.75$ (31 stars).  This 
needs to be borne in mind in the later discussion of results for Cyg~OB2.

\subsection{NGC 7790}

\begin{figure}
\begin{picture}(0,150)
\put(0,0)
{\includegraphics{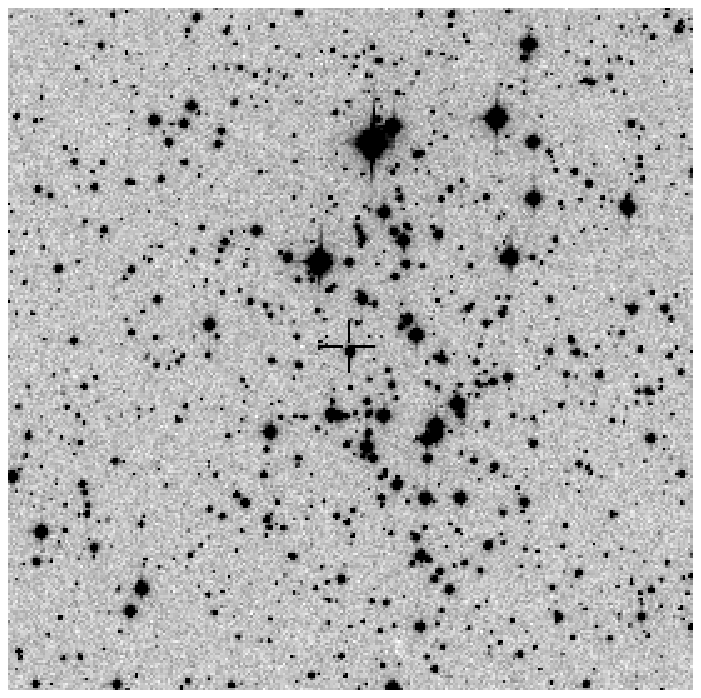}}
\end{picture}
\caption{The $6 \times 6$ arcmin$^2$ field centred on the test cluster 
NGC 7790, as extracted from the IPHAS database.  This is the 30-sec 
$r'$ exposures from pointing 7626o.  N is up and E to the left.
}
\label{fig_7790r}
\end{figure}

\begin{figure}
\begin{picture}(0,190)
\put(0,0)
{\includegraphics{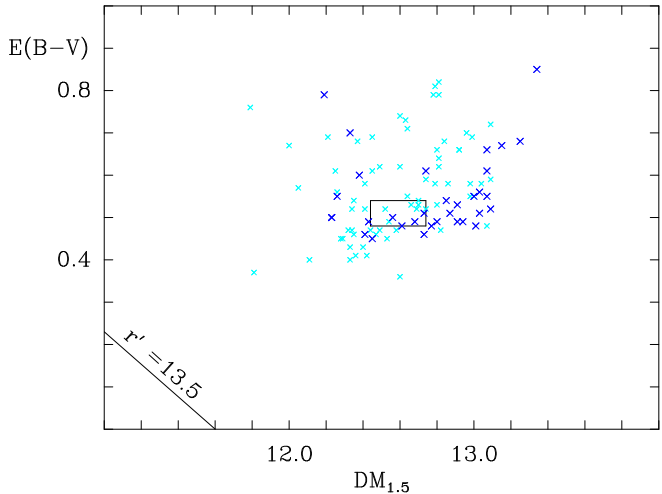}}
\end{picture}
\caption{Result for the open cluster NGC 7790.  Derived $E(B-V)$ and 
$DM_{1.5}$, are plotted against each other.  The larger sample shown in a 
lighter cyan colour 
was obtained from the $31\times31$~ arcmin$^2$ overlap of IPHAS fields 
7626/o, including NGC 7790.  The data shown in blue were obtained from the 
smaller $6\times6$~arcmin$^2$ field centred on the cluster.  The cut line in 
both cases was set at $\Delta (r'-H\alpha) = 0.025$.  The box, centred on 
$DM=12.59$ and $E(B-V)=0.51$, is the result for this cluster due to Gupta 
et al. (2000).}
\label{fig_ngc7790}
\end{figure}

The open cluster, NGC 7790, is believed to be about 100--120 Myrs old 
(Bragaglia \& Tosi 2006; Gupta et al. 2000) and hence close to the age at 
which A stars would begin leaving the main sequence.  We have extracted the 
photometry from a $6\times6$ 
arcmin$^2$ field centred at RA 23 58.4, Dec +61 13 (J2000, 
see Fig.~\ref{fig_7790r}) and have selected candidate A stars 
following exactly the same method as for NGC 7510. Given the age of this
cluster there is some room to question whether a more firmly main-sequence, 
slightly fainter absolute magnitude, $M(r') \simeq 1.9$, would work better
than the value of 1.5 more typical of the general field.  Staying with
$M(r') = 1.5$, we arrive at the result shown in Fig.~\ref{fig_ngc7790} as the 
blue datapoints.  The comparison is made in this case with the results of the 
photometric study by Gupta et al. (2000): they 
obtained $E(B-V) = 0.51\pm0.03$, $DM = 12.59\pm0.15$.

Once again, it is clear this method of A-star selection has found a group
of objects belonging to the target cluster at a range of reddenings compatible
with the results of detailed photometric study.  Had we adopted the fainter 
main sequence absolute magnitude the distribution of stars in 
Fig.~\ref{fig_ngc7790} would still have overlapped the Gupta et al distance,
but most of the A stars would have been lying at lower $DM$.  That the 
brighter field absolute magnitude does a little better suggests that some
stars have begun to evolve away from the main sequence.

A second extraction across a larger field of $31\time31$~arcmin$^2$, more than 
doubles the sample size and mildly smears the appearance of the $E(B-V)$ 
versus $DM_{1.5}$ diagram (the cyan data points in Fig.~\ref{fig_ngc7790}.  But
contamination of the sample by the general field appears slight.  This outcome 
may imply that A-star selection can become a useful tool for identifying new 
clusters with economy of effort.  This is the main way in which we use it 
below in the discussion of Cyg~OB2.

\section{Cyg OB2}
\label{cygob2}

\subsection{Early-A stars around the centre of Cyg OB2}

\begin{figure}
\begin{picture}(0,330)
\put(0,0)
{\includegraphics{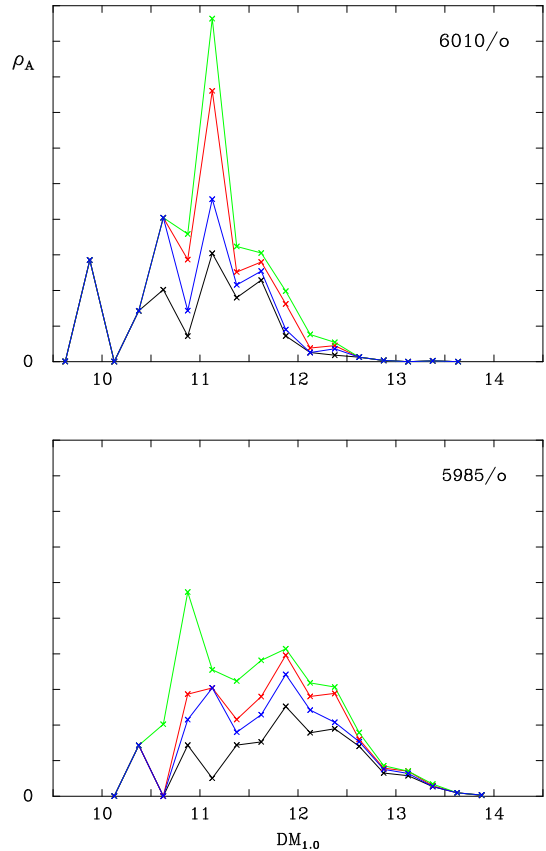}}
\end{picture}
\caption{The sampled variation of $\rho_A$ versus provisional distance modulus 
($DM_{1.0} \equiv$ dereddened $r'$ magnitudes less an adopted median absolute 
magnitude of 1.0) for fields 6010 and 5985 to either side of the centre 
of Cyg~OB2.  In each panel the four curves shown are for $\Delta(r'-H\alpha) = 
0.015$, 0.02, 0.025 and 0.03, and are again respectively black, blue, red and
green, as in Fig.~\ref{fig_control}.  The adopted bin size is 
$\Delta(DM) = 0.25$.  
}
\label{fig_hist}
\end{figure}

\begin{figure}
\begin{picture}(0,330)
\put(0,0)
{\includegraphics{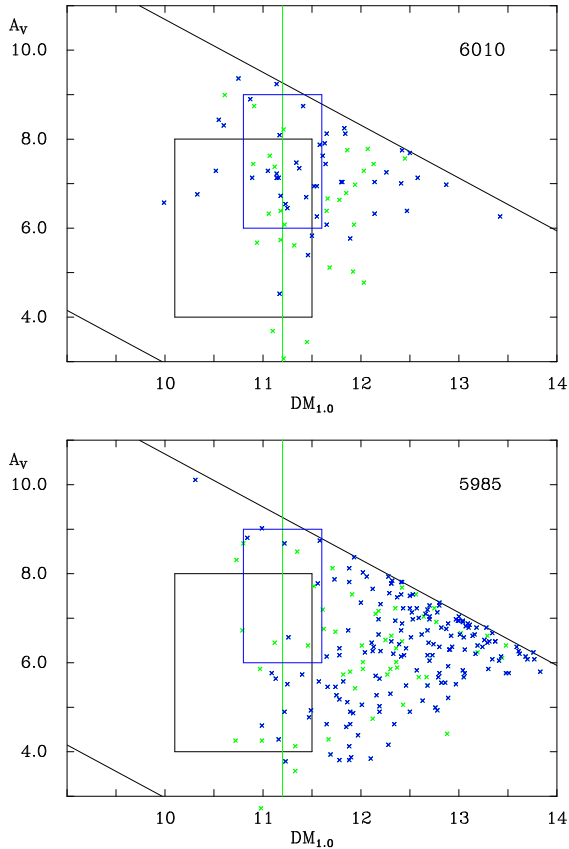}}
\end{picture}
\caption{Visual extinctions versus provisional distance moduli $DM_{1.0}$ 
for fields 6010 and 5985 straddling the centre of Cyg~OB2.  The dark 
blue points correspond to objects selected below the 
$\Delta (r'-H\alpha) = 0.02$ cut line (cf Fig.~\ref{fig_hist}), while 
the lighter green points are the objects added to the sample for the 
highest cut line ($\Delta(r'-H\alpha) = 0.03$).  The box drawn in black
and the green line are discussed in the text.  The box in blue delineates
a range in $DM_{1.0}$ also picked out in figs~\ref{fig_6001-5976} and 
\ref{fig_5998-5962}.}
\label{fig_DMvred}
\end{figure}

We now come on to the second main aim of this study: to access a lower 
stellar mass range than has hitherto been open to examination in 
Cyg~OB2 and its environment.   The reward for doing this will be, at
minimum, a better picture of the angular extent of this association.  

We first consider the selection and dereddening of candidate early-A dwarfs
in IPHAS fields 6010 and 5985, that straddle the centre of Cyg OB2
(Fig.~\ref{figure_fields}).  The colour-colour plane for field overlap
6010 was shown as Fig.~\ref{cc_data}, from which it was noted that 
Cyg OB2 is apparent as an excess of objects in the approximate reddening range 
$4.5 < A_V < 7$.  As in Fig.~\ref{cc_data}, no selected object will be 
fainter then $r' = 20$.  As for NGC 7510 and NGC 7790, this cut-off magnitude 
is 2 magnitudes brighter than the reduction pipeline estimate of the 
magnitude limit of the observations.  In this and all other respects, the 
method of selection is the same as those applied to these test-case clusters.

In Fig.~\ref{fig_hist}, we present the space density distributions for 
provisional distance moduli, $DM_{1.0}$, obtained from early-A dwarf 
candidates for different offsets of the IPHAS limit line for both fields: as 
the offset is increased, so the sample size increases.  We choose $M(r') = 
1.0$ for now, noting that this corresponds to an age in the region of 5~Myr.  
In its lowest position the cut line is placed at $\Delta(r'-H\alpha) = 0.015$ 
above the A0V reddening line.   The increasing contamination with increasing 
offset potentially brings in intrinsically redder, and fainter, stars, that 
might bias the derived distribution toward over-estimated reddenings and 
underestimated distance moduli.  A very small effect, possibly due to this, 
may be apparent in the populous, less reddened field (5985) to the NW of 
the nominal Cyg OB2 centre -- while the overall sample size grows from 142  
objects (black curve) to 223 (green curve).  In field 6010 the shift is 
negligible and perhaps just positive, while the sample size increases from 37 
objects (black curve) to 77 (green curve).  To limit Poissonian noise, we 
will prefer the largest samples obtained on setting 
$\Delta (r'-H\alpha) = 0.03$. 

The striking feature of Fig.~\ref{fig_hist} is the large difference in the
A-star density distributions contained in the two neighbouring fields.  
For 6010, the peak in the range $11.0 < DM_{1.0} < 11.25$ sits at 3-sigma
contrast with respect to the less well-populated bins to either side of it.
No such peak is apparent in 5985 and the overall population seems less
dense and more distant.  To better understand the origin of this, we show in 
Fig.~\ref{fig_DMvred}, the relationship between reddening ($A_V$) and 
$DM_{1.0}$ in both cases.  In field 6010 just to the SE of the nominal centre 
of Cyg OB2, it is evident that the stars making up the peak population around 
$DM_{1.0} \simeq 11$ are seen through visual extinctions of 5 to 8 
magnitudes.  This contrasts with 5985 where the most common reddening to 
$DM_{1.0} \sim 11$ is 4 to 6 magnitudes.   The other clear point of contrast 
is that it is the slower rise of reddening with distance within 5985/o that 
allows us sight, in this field, of so very many candidate A dwarfs at 
$DM_{1.0} > 12$.  We are seeing these objects through a relative reddening 
hole that has been known about since the work of Reddish et al. (1967, see
also Dickel \& Wendker 1978, Kn\"odlseder 2000).  
It is also evident (from Fig,~\ref{fig_DMvred}), that the $r' = 20$ magnitude 
limit applied slices through this more distant population.

The box drawn in black in both panels of Fig.~\ref{fig_DMvred} spans the range 
in distance modulus and extinction obtained by Hanson (2003) in her study of 
brighter OB stars.  The mean $DM$ she favoured was 10.8.  The vertical green 
line in both panels marks $DM = 11.2$, obtained by Massey \& Thompson (1991).  
The tidy picture seen in the case of NGC 7510, a relatively isolated young 
cluster, is not repeated here.  In the upper panel, for 6010, the Hanson 
(2003) box looks to be displaced to lower distance moduli than the selected
A-dwarfs: to bring the two into alignment it would be necessary to increase the
assumed A-dwarf absolute magnitude by $\sim$0.7.  The Massey \& Thompson 
(1991) distance fares
rather better in that the offset is only about 0.3 magnitudes.  With either
of these comparisons there is a problem in that increasing the adopted
absolute magnitude amounts to raising the age of the population, when 
existing estimates for the age of Cyg~OB2 are less than 5~Myr. 

In 5985 (lower panel of Fig.~\ref{fig_DMvred}), the Hanson (2003) 
Cyg~OB2 distance and reddening range exhibits even less connection with 
the apparent distribution of candidate early-A stars.  Much of the early-A 
population in this field is unlikely to belong to any cluster at a distance 
below $\sim$2~kpc ($DM_{1.0} \sim 11.5$).  Indeed, there may be a steady
rise in reddening with increasing distance modulus in this field, with little
imprint of Cyg~OB2 itself.

\subsection{Candidate A dwarfs across the wider Cyg OB2 region}

\begin{figure}
\begin{picture}(0,340)
\put(0,0)
{\includegraphics{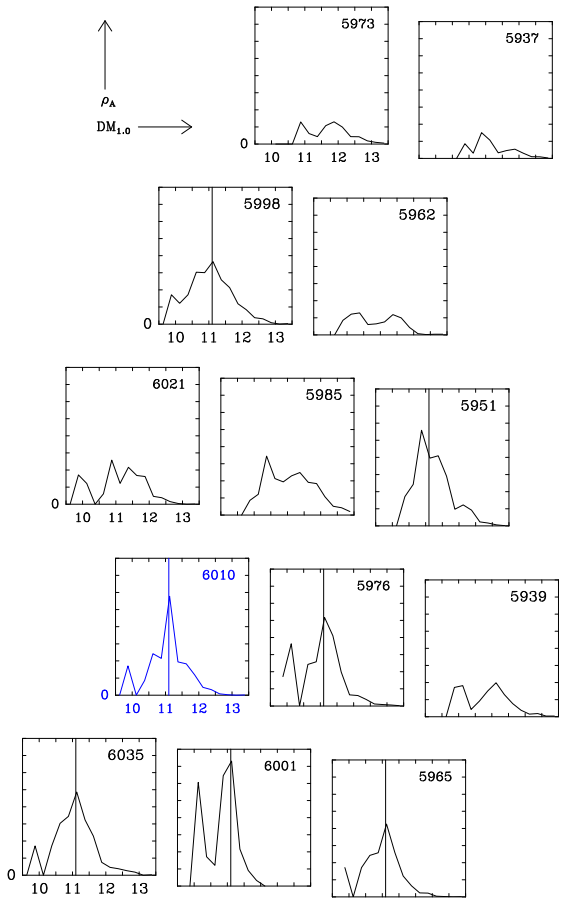}}
\end{picture}
\caption{Derived candidate early-A dwarf densities as a function of $DM_{1.0}$
for fields in and around Cyg OB2.  The relative positioning of the histograms 
roughly mimics their on-sky positioning, as shown in Fig.~\ref{figure_fields}.
Hence, roughly speaking,  N is up and E to the left.  To increase the sample 
sizes, the cut line in the IPHAS 
$(r'-H\alpha, r'-i')$ plane has been set at maximum 
($\Delta (r'-H\alpha = 0.03$).  The most nearly central field, 6010, is
picked out in blue.  Overlap fields listed in Table~\ref{field_list} 
with fewer than 50 candidate early-A dwarfs are not shown.  
In every panel, the axis ranges are the same and, in some, a vertical line 
is drawn at $DM_{1.0} = 11.1$.  
}
\label{fig_multi}
\end{figure}

The larger picture in this region is shown in Fig.~\ref{fig_multi}. 
Distributions are only shown for fields in which there are more than 50 
candidate early-A dwarfs (for $\Delta(r' - H\alpha) = 0.03$).  The number
of candidate objects per field was given in Table~\ref{field_list}.  The
better populated fields lie mainly in the centre and west of the region
studied. The outstanding feature of the A-star density profiles as a
function of $DM_{1.0}$, apparent from Fig.~\ref{fig_multi}, is the repeated 
density peak in the $11.0 < DM_{1.0} < 11.25$ bin in the southern fields 
(6010, 5976, 6035, 6001 and 5985).  High densities at similar 
distance moduli are also apparent in 5951 and 5998 -- but not in the 
two fields immediately to the north of the nominal Cyg~OB2 centre, 6021 
and 5985.

\begin{figure}
\begin{picture}(0,330)
\put(0,0)
{\includegraphics{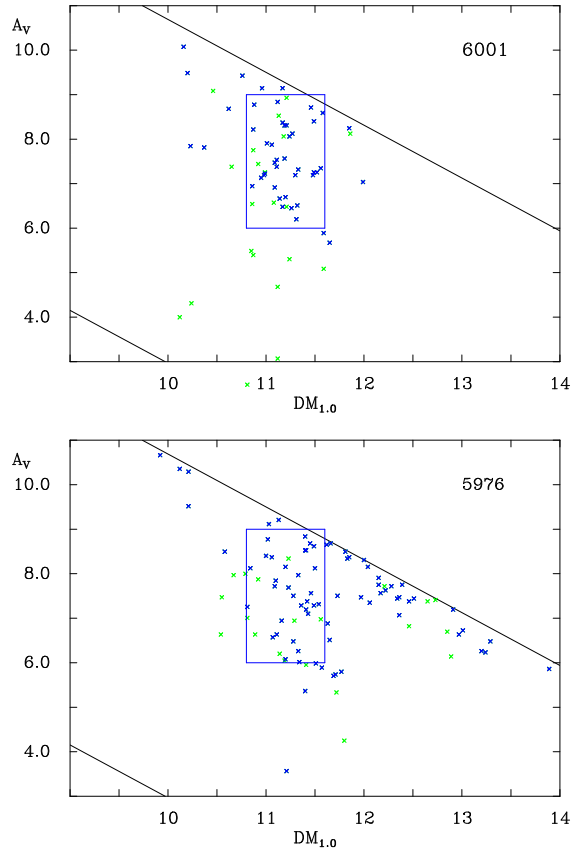}}
\end{picture}
\caption{Visual extinctions versus provisional distance moduli $DM_{1.0}$ 
for fields 6001 and 5976 to the south-west of the centre of Cyg~OB2.  
The dark blue points correspond to objects selected below the
$\Delta (r'-H\alpha) = 0.02$ cut line (cf Fig.~\ref{fig_hist}), while 
the lighter green points are the objects added to the sample for the 
highest cut line ($\Delta(r'-H\alpha) = 0.03$).  The $DM_{1.0}$
range picked out by the same blue box in each panel is the peak density 
region apparent from Fig.~\ref{fig_multi}.  To ease comparison, this 
region is also identified in both Figs.~\ref{fig_DMvred} and 
\ref{fig_5998-5962}.
}
\label{fig_6001-5976}
\end{figure}

To show in greater detail what this may mean, we plot reddenings against
$DM_{1.0}$ for fields 6001 and 5976 in Fig.~\ref{fig_6001-5976}.  The blue box 
is drawn around the main concentration of candidate early-A dwarfs found here 
($10.8 < DM_{1.0} < 11.6$).  Objects in this range make up the overwhelmingly 
dominant $\rho_A$ peak seen in Fig.~\ref{fig_multi} for field 6001 (39 stars, 
but note that the $\rho_A$ peak closer to $DM_{1.0} \sim 10$ is created by 
just 5 stars and so is much less significant).  The reddening in these fields 
is somewhat higher than in the central part of Cyg OB2 (6010) to the north: 
for most stars here, $A_V > 6$.  In field 5976 there is a tail of detected 
objects to much greater distances ($DM_{1.0} > 12$) which is absent from 6001, 
suggesting that the typical extinction affecting sightlines in 5976 
is less severe.  This also raises the possibility that the greater obscuration
in field 6001 induces more sample incompleteness: undercounting of A stars
in this field at $DM_{1.0} \gtrsim 11.0$ may be more significant.

\begin{figure}
\begin{picture}(0,330)
\put(0,0)
{\includegraphics{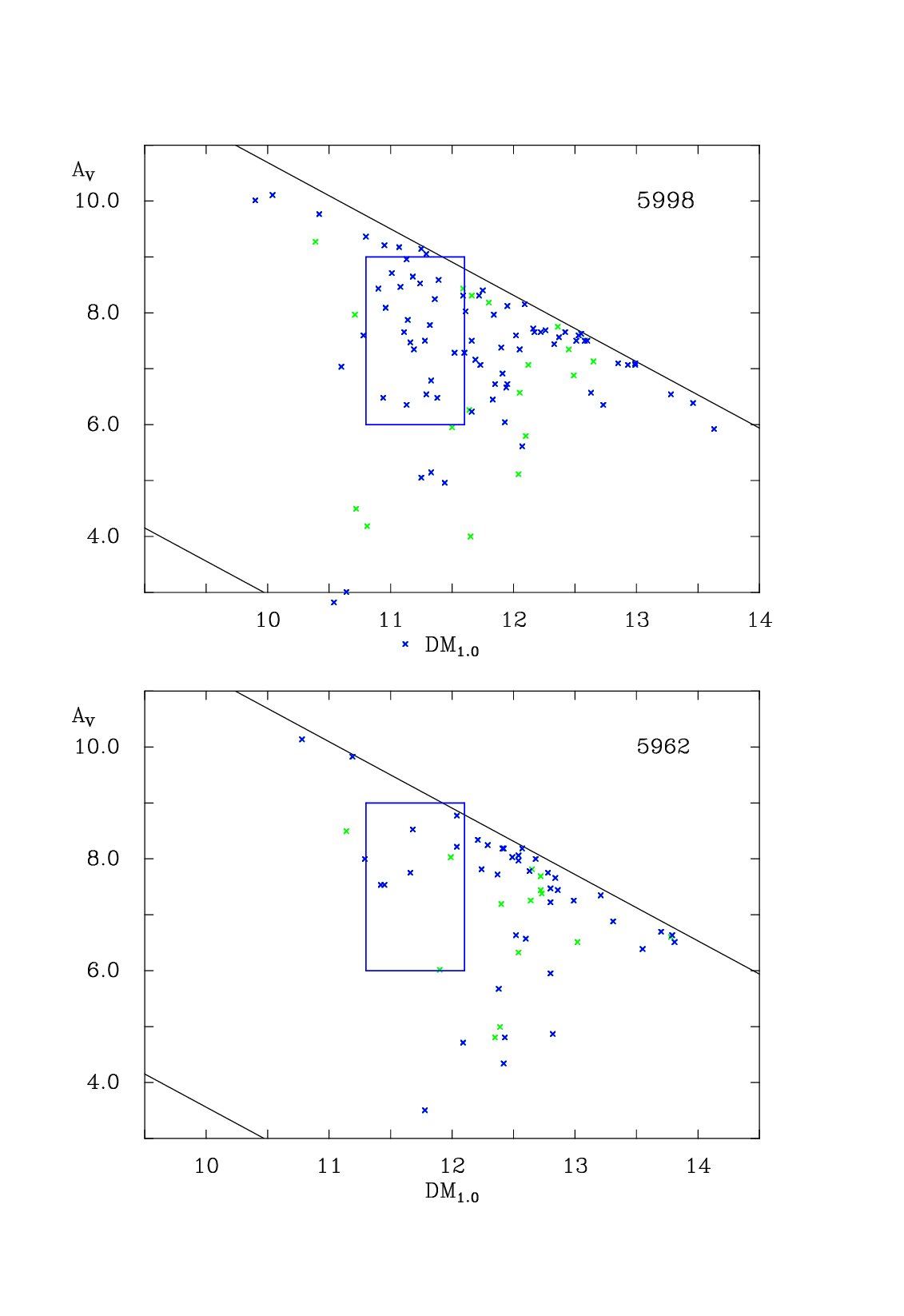}}
\end{picture}
\caption{Visual extinctions versus provisional distance moduli 
$DM_{1.0}$ for fields 5998 and 5962 to the north and north-west 
of the centre of Cyg~OB2.  The colours have the same significance as 
in Figs.~\ref{fig_DMvred} and \ref{fig_6001-5976}.
}
\label{fig_5998-5962}
\end{figure}
 
Finally, we take a look at 0.5--1 degree N and NW of the centre of Cyg~OB2, 
at fields 5998 and 5962 (Fig.~\ref{fig_5998-5962}).  Field 5998/o is 
rather similar to 6010, and fields further south.  The character of 5962 
is quite different: there is not a concentration of stars in the $DM_{1.0}$
range, 11.0 -- 11.5, and the overall number of candidate A dwarfs is relatively
low.  The modest pile up at $A_V \sim 8$ at distance moduli 12.0 to 13.0 in 
5962 (Fig.~\ref{fig_5998-5962}) suggests there may be more undiscovered at 
still higher reddenings, fainter than the imposed $r' = 20$ limit.  Inspection 
of similar reddening plots (not shown) for the fields neighbouring 
5962 supports the impression that incompleteness of this kind is much more of 
an issue in the extreme north of the region.

\subsection{Implications for the extent and nature of Cyg~OB2}

\begin{figure*}
\begin{picture}(0,340)
\put(0,0)
{\includegraphics{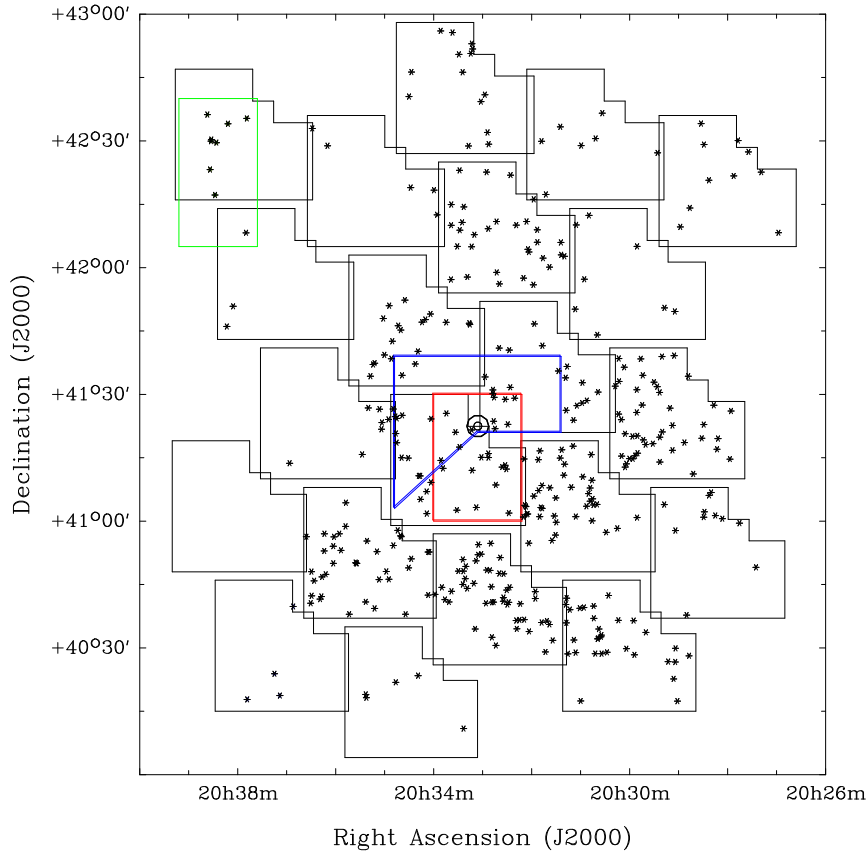}}
\end{picture}
\caption{The spatial distribution of candidate A dwarfs in the vicinity
of Cyg OB2, with dereddened $r'$ magnitudes in the range 
$11.75 \leq r'_0 \leq 12.5$.  The area bounded by the red line roughly 
matches the area of peak NIR stellar density as identified by Kn\"odlseder
(2000).  The area bound by the blue line picks out the main concentration
of optically identified OB stars (Massey \& Thompson 1991).  In the NE the
green box encloses the stellar population associated with DR 21 discussed
by Kumar et al. (2007).
}
\label{fig_map}
\end{figure*}

The main implication of Fig.~\ref{fig_multi} is the clear A-star
density peak centred on $11.0 < DM_{1.0} < 11.25$ in the southern fields,
that is most pronounced in field 6001.  Is this the real Cyg OB2?
This can be looked at in a different way by mapping the positions
of the stars with dereddened $r'$ magnitudes falling in the range 
$11.75 < r'_0 < 12.5$ that contribute to this apparent 'clustering' 
(Fig.~\ref{fig_map}).

On the basis of 2MASS data, Kn\"odlseder (2000) demonstrated a peak in 
stellar density at NIR wavelengths that coincides with the heart of
Cyg OB2 itself.  This region is enclosed by the the red box drawn in 
Fig.~\ref{fig_map}.  There is not a matching peak in the concentration of 
A stars here: the IPHAS field pair 6010 does not stand out in this way.  
Moreover, there is a relatively empty region cutting across the NE corner 
of 6010, continuing through the south of adjacent fields 6021 and 
5985.  This underpopulated region is much the same as that occupied by 
the OB stars picked out in Massey \& Thompson's (1991) optical study (the 
rough outline to this region is picked out in blue in Fig.~\ref{fig_map}, 
see also Fig. 4 of Hanson 2003).  Inspection of the original IPHAS
imaging confirms that this anti-correlation is not due to the OB stars 
creating a bright background swamping the fainter A stars. 

As shown in Fig.~\ref{fig_map} the east of the region is even more striking 
for the lack of candidate A dwarfs.  This absence was already foretold by 
the data in Table~\ref{field_list}, showing that there are few ordinary 
A stars to be had at any distance modulus.  The appearance of a few in the 
extreme north-east, in the vicinity of DR~21 (marked in green), may signal 
the emergence of the new generation of stars linked to this well-known HII 
region.  The median $r'_0$ for the group is 12.33.  If indeed they are as 
young (1--3~Myrs) as the population of objects analysed by Kumar et al.(2007), 
they would be 2.9~kpc away (adopting $M(r')=0$) -- the distance also
favoured by Kumar et al. (2007).

The major concentration of A dwarf candidates in Fig.~\ref{fig_map} is in
the southern fields that also stood out in Fig.~\ref{fig_multi}, albeit with
a zone of avoidance in the south west (lower part of 5976, extending to 
most of 5939 further west).  If the central field, 6010, is included with 
these fields (5951, 5965, 5976, 6001, 6035), the total number of 
plotted objects is 239 -- an average of 40 per field.  6001 is the most 
populous with 49 stars, 6010 the least with 32.  This leaves just 
125 objects to be shared among the remaining 14 fields in this magnitude
range.  In broad brush terms, we have roughly 200 A stars spread across an 
area about 1 degree across, distributed around RA 20h 32m and Dec $+$41 00 
(2000).

The typical dereddened magnitude for this grouping is $r'_0 \simeq 12.1$.
To be sure of the distance to this structure, an age must be known.  The
assumption made for Fig.~\ref{fig_multi} was consistent with an age of 
5 Myr, and placed this clustering at distance of 1.7~kpc ($DM_{1.0} \simeq 
11.1$).  If the typical age can be allowed to be as low as 2~Myr, then 
$M(r')$ is required to be as bright as 0.0, driving the distance up to an
improbable 2.6~kpc ($DM_{0.0} = 12.1$).   A further argument against this 
solution is that we are seeing evidence here that such young A star 
populations tend to be sparse -- as illustrated by the low counts linked to 
both DR~21 in the NE and to the main central concentration of 
optically-detected OB stars -- both likely to be only 1--3~Myrs old. 
It is more plausible to consider ages greater than 5~Myrs, rather than less.  
For example, an age of $\sim$7~Myrs imposes $M(r') = 1.4$ (Table~\ref{tab_Mr}) 
which then points to a distance modulus close to the value of 10.8, 
equivalent to a distance of 1.4~kpc, that was favoured by Hanson 2003 
based on revised OB-star calibrations. 

Accepting an age spread in and around Cyg~OB2 seems the best option.
If there are very young A-type objects embedded among the OB stars, 
they are few in number and/or escaping selection by 
virtue of either modest H$\alpha$ emission or high extinction ($A_V > 10$).   
The larger numbers of A stars to the south of Cyg OB2 centre give away the 
presence of the larger, somewhat older, main cluster.  In effect Cyg OB2 can 
be seen as a further example of cluster 'rejuvenation', but on a much 
grander scale than that already noted in the test case of NGC 7510.  What is
unclear at the moment is whether the rejuvenation is on the NE periphery of 
the main cluster, or whether it is more centrally located within a cluster 
that extends as far north as field 5998, which also exhibits an A-star 
density peak in the appropriate dereddened magnitude range 
(Fig.~\ref{fig_multi}).  The less elegant solution at this point is to 
view the swathe of A stars to the south of the nominal centre of Cyg OB2 
as evidence of a distinct cluster, of unknown age and distance, unconnected
with the known OB stars.  

If we can associate $\sim$200 A0-A5 near main sequence stars with a 5-7~Myr old
southerly cluster, then the combination of Siess et al. (2000) isochrones 
for this age range and the Kroupa (2001) flattening of the Salpeter IMF
below 0.5~M$_{\odot}$ would imply a total cluster mass in the region of 
10,000 M$_{\odot}$, and an OB star count of around 200.  There are a number 
of respects in which this estimate is surely on the low side.  First, a minor 
point is that the sky coverage is only 90 percent.  But more important is the 
fact that the selection of A stars applied here has been fairly conservative 
in order to keep sample contamination minimal: had the cut 
line been raised to $\Delta(r' - H\alpha) = 0.05$, the number of candidate 
A dwarfs in the region would nearly double.  There is also clear 
evidence of sample incompleteness due to high and patchy reddening -- a 
problem that appears more marked in the more southerly fields 6001 and 5976
with higher $\rho_A$ than near the OB star concentration.  And 
finally, this accounting does not include the immediate environs of the
OB stars in which fewer stars of early-A colour have yet appeared: applying 
the same approach as above for a star count of $\sim$30 (for e.g. field 6010) 
at the much younger age of $\sim$2~Myrs adds a further 
$\sim$10,000~M$_{\odot}$.  A final proper accounting of the total mass in the 
region of 30,000 to 40,000 M$_{\odot}$ would not be surprising.

The reputation of Cyg~OB2 as the outstanding visible northern OB 
association remains untarnished.  However, the upper limit
of the claim by Kn\"odlseder (2000) is difficult to sustain:  if 
this region truly is home to over 2000 OB stars then this appraisal 
of the A star content has under-counted by an order of magnitude.
At least in the core region of Cyg OB2 where the reddening is less
severe, so large a discrepancy is unlikely. 

\section{Closing Remarks}

This study has had two goals.  The first of these was to show how the IPHAS 
$(r'-H\alpha, r'-i')$ plane may be exploited in a very simple way to
yield reliable samples of near main-sequence A0-A5 stars.  Such objects 
in the general field, selected as described here and validated by HIPPARCOS
data, will present themselves with a typical absolute magnitude 
$M(r')$ of 1.5 -- 1.6.  Fortunately, for stellar ages of $\sim$ 10~Myrs or 
more there is little
change in this absolute magnitude.  For simple single-aged populations a 
magnitude spread of $FWHM \sim 0.5$ is to be expected.  Demonstrations of 
this selection at work have been provided by the examples of the open 
clusters NGC 7510 and NGC 7790.

The second aim has been to apply this A-star selection technique to the
challenging example of Cyg~OB2.  This spectacular northern cluster, 
immersed in the Cygnus X region of the Galactic Plane, continues to be
debated in the literature on account of its uncertain dimensions and
distance.  The new capability directly bestowed by IPHAS photometry
(unsupported, thus far, by spectroscopy) is that of identifying the
on-sky distribution of and reddenings of over 1000 A0-A5 stars towards, 
around, within, and beyond Cyg OB2.  The result of this is a clear peak in 
A-star density, comprising $\sim$200 objects, roughly centred on RA 20h 32m 
and Dec $+$41 00 (2000), some 20 arcmin south of the VI~Cyg No.8 trapezium 
normally taken as the centre of Cyg~OB2.

The most parsimonious interpretation of the results obtained is to treat
the southern clustering of A stars and the already known OB star 
concentration as part of the same cluster, compatible with a distance
of 1.4~kpc ($DM \simeq 10.8$, Hanson 2003) if the A stars are $\sim$7~Myr 
old, or with
1.7~kpc ($DM = 11.2$, Massey \& Thompson 1991) if their typical age is
$\sim$5~Myr.  Either solution requires some age contrast between the A stars 
(5-7 Myrs old) and the main group of OB stars (1--3 Myrs old, Hanson 2003).  
This may amount to two main bursts of star formation, or to an indication of 
a substantial age spread (cf. the case of the ONC cluster discussed by Palla 
et al. 2005).  The presence of a range of ages may help to explain why  
Cyg~OB2 is not associated with bright H~{\sc ii} emission despite its very 
evident rich O star population.  The massive cluster, R~136 in the Large 
Magellanic Cloud exhibits similar behaviour in that its intermediate mass stars
appear to be $\sim$6~Myrs old, as compared with 1--2~Myrs for the most massive
stars (see Massey 2003). Indeed, in Cyg~OB2 itself, signs of non-coeval star 
formation have been noticed before (Massey \& Thompson 1991, Hanson 2003).  
The mass linked to the newly-found A stars in the cluster is at least 
10,000~M$_{\odot}$.  As 
a cautionary note, we point out that it remains possible (if less attractive, 
see Schneider et al 2006) that the main A-star clustering is at a different 
distance from the OB stars.

There are two wider applications of this new technique for selecting A 
stars from IPHAS.  First, it is a new tool for use in exploring the 
structure of the Galactic thin disk.  This will be especially incisive
when applied to the less reddened outer Galaxy.  We have not discussed 
metallicity here, but it should not create much difficulty: 
on the one hand, the weak line-blanketing at red wavelengths implies 
that the $r'-i'$ colour of A stars will not be metallicity-sensitive, and, 
on the other, the evidence to date is that metallicities ($[Fe/H]$) in the 
thin disk at large Galactic radii are not greatly subsolar (e.g. Carraro et 
al 2007, who find a mean $[Fe/H]$ of $\sim -0.35$ for $R_G$ between 12 and 
21~kpc).  Second, A-star selection over large sky areas allows a broad census 
to be conducted of the early evolution of A stars by, for example, 
identifying the fraction with clear infrared excesses.  A study of this 
kind has already been undertaken by Hales et al. (2008), who have combined 
IPHAS photometry with Spitzer/GLIMPSE and 2MASS data.  

A work for the future is to generalise the A-star selection scheme to begin
to extract a broader range of spectral types from the main stellar locus in
the IPHAS colour-colour plane.  In principle it is possible to map the
positions of A-K near main-sequence stars as a function of reddening onto 
unique $(r'-H\alpha, r-i')$ values -- if an effective means to
distnguish and separate out higher luminosity classes can be identified.  
An exploration of this opportunity is now underway (Sale et al., in 
preparation). 

\section*{Acknowledgments}
We thank Kerttu Viironen who was the observer at the telescope the
night the Cyg~OB2 data were obtained.  We also thank Jorick Vink, 
Danny Steeghs, Danny Lennon and Jeremy Drake for helpful comments relating to 
this work.  The referee of this paper, Philip Massey, is thanked for his 
useful comments.  This paper makes use of data obtained as part of the
INT Photometric H$\alpha$ Survey of the Northern Galactic Plane (IPHAS) carried
out at the Isaac Newton Telescope (INT). The INT is operated on the island of 
La Palma by the Isaac Newton Group in the Spanish Observatorio del Roque 
de los Muchachos of the Instituto de Astrofisica de Canarias. All IPHAS data 
are processed by the Cambridge Astronomical Survey Unit, at the Institute of 
Astronomy in Cambridge. We also acknowledge use of data products from the 
Two Micron All Sky Survey (2MASS), which is a joint project of the University 
of Massachusetts and the Infrared Processing and Analysis Center/California 
Institute of Technology (funded by the USA's National Aeronautics and Space 
Administration and National Science Foundation).  Stuart Sale is in receipt 
of a studentship funded by the Science \& Technology Facilities Council of 
the United Kingdom.

\end{document}